
\magnification=1200
\documentstyle{amsppt}
\NoBlackBoxes
\define\sel{\operatorname{sl}}

\topmatter
\title
$q$--Hypergeometric Functions in the Formalism of Free Fields
\endtitle
\author
Alexei Morozov*\\ Luc Vinet**
\endauthor
\leftheadtext{Alexei Morozov and Luc Vinet}
\affil
*permanent address: ITEP, 117259, Moscow, Russia
\medskip
**Centre de Recherches Math\'ematiques, Universit\'e de Montr\'eal, C.P. 6128
-- A, Montr\'eal (Qu\'ebec) Canada H3C 3J7
\bigskip\medskip
\centerline{CRM-1884/ITEP-M8}
\endaffil
\abstract
We describe a representation of the $q$--hypergeometric functions of one
variable in terms of correlators of vertex operators made out of free scalar
fields on the Riemann sphere.
\endabstract
\endtopmatter

\document
\head 1. Introduction \endhead

The $q$--hypergeometric functions \cite{1} are of great interest in modern
mathematical physics because of the clues that they are expected to give in
connection with the development of the theory of difference equations and of
quantum and non-commutative geometry.  The main idea is to consider the
$q$--hypergeometric functions as associated with the quantum analog of the
Riemann sphere, which in turn is supposed to provide an interpolation between
the Riemann sphere itself, its non-compact analogue --- the upper half-plane
--- and the $p$--adic counterparts of the latter, represented for example, by
discrete spaces like Bruhat Tits trees.  Among other things, it is hoped that
this approach will lead to a better understanding of the geometry of the
simplest quantum group $Sl(2)_q$.  It could also suggest interesting
generalizations of $q$--hypergeometric functions which could reflect the
properties of both the generic quantum groups and the quantum analogs of spaces
of arbitrary topology.  An immediate appli

cation would be a theory of integrable hierarchies, containing both
differential and difference equations, i.e. both the KP and Toda-like systems.
One could hope to get a deeper understanding of old powerful techniques like
the various versions of the Bethe ansatz, the Yang--Baxter equations and the
theory of lattice integrable systems like $XYZ$--model, as well as to establish
their explicit relation to the theory of KP and Toda hierarchies.  Two
different aspects of this promising program have been recently discussed in
some details in refs. \cite{2} and \cite{3}.

The purpose of this letter is much more modest: it is to attract attention to
the possible role of the free field formalism in the future development of
these ideas.  The free 2--dimensional massless holomorphic quantum fields have
their pair correlators given by
$$
\langle \phi^\mu(t) \phi^\nu(t^\prime) \rangle = \delta^{\mu \nu} \log(t -
t^\prime) + \text{regular terms} \tag1
$$
and all other correlators estimated by the Wick rule.\footnote{A short
terminological comment may be in order.  A field is called {\it free} whenever
its correlators obey the Wick rule.  We use the word ``massless'' for
two-dimensional scalar fields on the Riemann sphere, whenever the pair
correlators are (linear combinations of) logarithms.  Note that $2d$ Lorentz
invariance is not required.  This definition allows to consider not only
``holomorphic'' fields, as in eq. (1), but also somewhat more sophisticated
free fields like those of (44).} They are known to play a central role in all
the theories listed above as subjects to be unified by the theory of
$q$--special functions.  Indeed, string models with the Riemann sphere, the
upper half-plane or Bruhat-Tits trees as world sheets, are usually described in
terms of free fields (see \cite{4} for the least known case of $p$--adic
strings).  Quantum groups arise naturally in the study of rational conformal
theories (see the already cited review \cite{2} and r

eferences therein).  These in turn, are describable in terms of free fields
(i.e. in the Feigin--Fucks \cite{5} or Dotsenko--Fateev \cite{6} formalism),
either through the minimal models \cite{7} or the Wess--Zumino--Novikov--Witten
model and its reductions \cite{8}.  We note that the last model is intimately
related to the coadjoint-orbit approach in group theory \cite{9}.  As to the
integrable hierarchies, they are identified with the theory of free fields
through the concepts of infinite Grassmannian and $\tau$--function \cite{10},
which are further related to the theory of random matrices and orthogonal
polynomials (see \cite{11} for a review).  Integrability appears also reflected
in the topological properties of the moduli spaces \cite{12} and this brings us
back to the starting point: strings living on various Riemann surfaces.  It
should be mentioned that there has been recently a lot of interest
\cite{13--17} in free field realizations of quantum (affine) algebras as tools
for solving the $q$--Knizh

nik--Zamolodchikov equations that are obeyed for instance by the correlation
functions of the $XXZ$ spin chain \cite{18,19}.

While in the general context of string theory (and integrable hierarchies), the
free fields with arbitrary boundary conditions are important, there are special
cases when the ``simplest'' free fields --- those on the Riemann sphere --- are
of interest.  In particular, this could be a nice place from where to begin the
study of quantum geometry.  In this case formula (1) is exact: no ``regular
terms'' appear on the r.h.s.  It is in this situation that $q$--{\it
hypergeometric} functions arise and we shall here restrict our considerations
to this case.

The ordinary $(q=1)$ hypergeometric functions can be represented as correlators
of the ``spherical'' free fields, or, more exactly, of the ``vertex operators''
$V_\alpha (z) = e^{i \alpha \phi(z)}$: (here $\alpha \phi = \sum_\mu \alpha^\mu
\phi^\mu$) and the ``screening charges'' $Q_{\gamma, C} = \oint_C V_\gamma (t)
dt$.\footnote{This construction can be related to the ``orbit construction'',
where the same functions are represented as the matrix elements $\langle M
\vert e^{\beta_+J_+}e^{\beta_-J_-}\vert N \rangle$, with $J_\pm$ the raising
and lowering operators associated with the positive and negative roots of
certain Lie groups, and with $\langle M \vert \ \text{and}\ \vert N \rangle$
belonging to some representation space of the group.  In this approach the
screening charges are essentially included in the definition of $\langle M
\vert \ \text{and}\ \vert N \rangle$ (see \cite{20} for some details).  For the
$q$--generalization of this ``orbit construction'' see \cite{21}.}  We shall
now present this

 construction in a form that will have a clear generalization to the
$q$--hypergeometric case.

\head 2. Integral --- and free field representations of ordinary hypergeometric
functions \endhead

The ordinary hypergeometric functions of one variable are defined by the
following series
$$
{}_rF_s(a_1 , \dotsc , a_r; b_1 , \dotsc , b_s; z) \equiv \sum_{n=0}^\infty
\frac{(a_1)_n \dotsb (a_r)_n}{(b_1)_n \dotsb (b_s)_n} \frac{z^n}{n!} \tag 2
$$
where
$$
(a)_n \equiv \frac{\Gamma(a+n)}{\Gamma(a)} = a(a+1) \dotsb (a+n-1). \tag 3
$$
Among these functions, two are elementary
$$
\gather
{}_0F_0(z) = e^z \tag4\\
{}_1F_0(a;z) = \sum_{n=0}^\infty \frac{\Gamma(a+n)}{\Gamma(a)n!}{z^n} =
\sum_{n=0}^\infty \frac{\Gamma(1-a)}{\Gamma(1-a-n)n!}{(-z)^n} = (1-z)^{-a}
\tag5
\endgather
$$
while the others are in general transcendental: ${}_0F_1(b;z)$ is for instance
related to Bessel functions.  Either one of the functions ${}_0F_0(z)$ or
${}_1F_0(a;z)$ can be used as starting point for a recursive construction of an
integral representation of the functions ${}_rF_s$.  This construction involves
three elementary steps:
$$
\align
{}_rF_s
&\longrightarrow {}_{r+1} F_{s+1} \tag6\\
{}_{r+1}F_s
&\longrightarrow {}_rF_s \tag7\\
{}_rF_{s+1}
&\longrightarrow {}_rF_s. \tag8
\endalign
$$
By combining these one clearly can transform any ${}_r F_s$ into any other
${}_{r^\prime}F_{s^\prime}$.  The operations (6)--(8) are explicitly realized
as follows.

\demo{Step (a)}
$$
\multline
{}_{r+1} F_{s+1} (a_1 , \dotsc , a_{r+1}; b_1, \dotsc , b_{s+1}; z)=\\
\frac{1}{\widehat B(a_{r+1}, b_{s+1})} \int_0^1  dt t^{a_{r+1}-1}
(1-t)^{b_{s+1}-a_{r+1}-1} {}_rF_s(a_1, \dotsc , a_r; b_1, \dotsc , b_s; tz).
\endmultline \tag9
$$
This identity (which was discussed long ago \cite{22})  is a simple corollary
of the definition (2) and of the integral representation of the beta function:
$$
\widehat B(a,b) \equiv \int_0^1  dt \,t^{a-1}(1-t)^{b-a-1} = \frac{\Gamma(a)
\Gamma(b-a)}{\Gamma(b)}. \tag10
$$
We are using the unconventional notation $\widehat B(a,b) \equiv B(a,b-a)$ to
simplify the formulas.  Indeed, what is needed is the linear operation $t^n
\longrightarrow (a_{r+1})_n /(b_{s+1})_n$ and this is provided by
$$
\multline
\int_0^1 dt t^{a_{r+1}+n-1} (1-t)^{b_{s+1}-a_{r+1}-1} = \\
\frac{\Gamma(a_{r+1}+n) \Gamma(b_{s+1} - a_{r+1})}{\Gamma(b_{s+1}+n)} =
\frac{(a_{r+1})_n}{(b_{s+1})_n} \widehat B(a_{r+1}, b_{s+1})
\endmultline \tag11
$$
\enddemo

The steps (7) and (8) are easily made explicit from observing that
$\botsmash{\lim\limits_{N \to \infty}} (N)_n/N^n \allowmathbreak =~1$.  From
this we get
$$
\aligned
&{}_rF_s(a_1, \dotsc , a_r; b_1, \dotsc , b_s;z)\\
&\quad=\lim_{N \to \infty} {}_{r+1}F_s(a_1, \dotsc , a_r, a_{r+1} = N; b_1 ,
\dotsc , b_s; z/N)\\
&\quad= \lim_{N \to \infty} {}_rF_{s+1} (a_1, \dotsc , a_r; b_1 , \dotsc , b_s,
b_{s+1} = N; Nz).
\endaligned \tag 12
$$
For our purposes, it will suffice to interpret all these formulas as relations
between formal series.

Although (6--7) (or (9)--(12)) are enough to reproduce from ${}_0F_0$ or
${}_1F_0$ the entire set of hypergeometric functions, there exist different
ways of realizing the steps (6)--(8).  For example:  ${}_rF_s \to {}_{r+1}F_s$
(i.e. $t^n \to (a_{r+1})_n$) can be obtained from
$$
\multline
{}_{r+1}F_s(a_1, \dotsc , a_{r+1}; b_1, \dotsc , b_s; z)\\
= \frac{1}{\Gamma(a_{r+1})} \int_0^\infty dt e^{-t} t^{a_{r+1}} {}_rF_s(a_1,
\dotsc , a_r; b_1, \dotsc , b_s; tz)
\endmultline \tag13
$$
while ${}_rF_s \to {}_rF_{s+1}$ (i.e. $t^{-n} \to 1/(b_{s+1})_n$) can be gotten
from
$$
\multline
{}_rF_{s+1}(a_1, \dotsc , a_r; b_1, \dotsc , b_{s+1}; z)\\
= \frac{1}{\Gamma(1-b_{s+1})} \int_0^\infty dt e^{-t} t^{-b_{s+1}} {}_rF_s(a_1,
\dotsc , a_r; b_1, \dotsc , b_s; z/t).
\endmultline \tag14
$$
Let us also present two other versions of (9):
$$
\split
{}_{r+1}&F_{s+1} (a_1, \dotsc , a_{r+1}; b_1, \dotsc , b_{s+1};z)\\ &=
\frac{1}{\widehat B(a_{r+1}, b_{s+1})} \int_1^\infty  dt\, t^{-b_{s+1}}
(t-1)^{b_{s+1}-a_{r+1}-1} {}_rF_s(a_1, \dotsc , a_r; b_1, \dotsc b_s; z/t)\\ &=
\frac{z^{-a_{r+1}}}{\widehat B(a_{r+1},b_{s+1})} \int_0^z  dt\, t^{a_{r+1}-1}
(1 - t/z)^{b_{s+1} - a_{r+1} - 1} {}_rF_s (a_1, \dotsc , a_r; b_1, \dotsc b_s;
t),
\endsplit \tag 15
$$
We now return to free fields and note that
$$
\langle V_{\vec\alpha} (z) V_{\vec{\alpha}^\prime}(z^\prime)\rangle =
\langle e^{i\vec\alpha \cdot \vec\phi(z)} e^{i \vec{\alpha^\prime} \cdot
\vec\phi(z^\prime)} \rangle = (z-z^\prime)^{-\vec\alpha \cdot
\vec\alpha^\prime}.
\tag 16
$$
In particular
$$
\langle e^{i\vec\alpha_1 \phi(1)}e^{i \vec\alpha_z \phi(z)} \rangle =
(1-z)^{-\vec\alpha_1 \cdot \vec\alpha_z} = {}_1F_0(\vec\alpha_1 \cdot
\vec\alpha_z; z). \tag17
$$
In conjunction with formulas (11) and (12), this means that all the functions
${}_{s+1}F_s$ (i.e. those with $r-s=1$) can be immediately represented as
integrals of correlators of the vertex operators, while all the other ${}_rF_s$
(with $r-s \ne 1$) can be obtained as limits of these ${}_{s+1}F_s$.  One might
note that precisely those $(q)$--hypergeometric functions with $r-s=1$ seem to
have the most interesting applications.  Whether this is fortuitous or has
something to do with their more natural relation with the free field formalism
is an interesting question.  Explicitly, the representation of ${}_{s+1}F_s$ is
as follows:
$$
\split
&{}_{s+1}F_s (a_1, \dotsc , a_{s+1}; b_1, \dotsc , b_s; z)=
\frac{z^{1-b_s}}{\widehat B(a_{s+1},b_s)} \int_0^z dt_s t_s^{a_{s+1}-1} (z -
t_s)^{b_s - a_{s+1} -1} \\
&\qquad\times \frac{t_s^{1-b_{s-1}}}{\widehat B(a_s, b_{s-1})} \int_0^{t_s}
dt_{s-1} t_{s-1}^{a_{s-1}-1} (t_s - t_{s-1})^{b_{s-1} - a_{s}-1}\\
&\qquad\dotsb\times
\frac{t_2^{1-b_1}}{\widehat B(a_2, b_1)} \int_0^{t_2}  dt_1
t_1^{a_2-1}(t_2-t_1)^{b_1-a_2-1}(1-t_1)^{-a_1}\\
&= z^{1-b_s} \prod_{j=1}^s \left( \int_0^{t_{j+1}}  dt_j
\frac{t_j^{a_{j+1}-b_{j-1}}(t_{j+1} - t_j)^{b_j -a_{j+1}-1}}{\widehat
B(a_{j+1}, b_j)} \right) t_1^{b_{j-1}} (1-t_1)^{-a_j}\\
&= z^{-a_{s+1}} \prod_{j=1}^s \left( \int_0^{t_{j+1}}  dt_j
\,t_j^{a_{j+1}-a_j-1}
(1 - t_j/t_{j+1})^{b_j-a_{j+1}-1}\right) t_1^{a_1} (1-t_1)^{-a_1}\\
&= \left(\prod_{j=1}^s \frac{1}{\widehat B(a_{j+1}, b_j)} \right)\langle
V_{\vec\alpha_z} (z) V_{\vec\alpha_0}(0) V_{\vec\alpha_1}(1) \prod_{j=1}^s
\int_0^{t_{j+1}} dt_j V_{\vec\gamma_j}(t_j) \rangle.
\endsplit\tag18
$$
We have put $t_{s+1} \equiv z$.  It is convenient to set $t_0 = 0$, $t_{s+2} =
1$ and $\vec\alpha_0 = \vec\gamma_0$, $\vec\alpha_z = \vec\gamma_{s+1}$,
$\vec\alpha_1 = \vec\gamma_{s+2}$.  The $\vec\gamma_s$ should then be chosen so
that
$$
\alignedat2
\vec\gamma_i \cdot \vec\gamma_j
&= (a_j - b_i + 1) \delta_{j,i+1} \quad &&\text{for} \quad 1 \le i < j \le
s+1\\
\vec\gamma_j \cdot \vec\gamma_{s+2}
&= a_1 \delta_{j_1} \quad &&\text{for} \quad 1 \le j \le s+1\\
\vec \gamma_1 \cdot \vec\gamma_0
&=1-a_2 && \\
\vec\gamma_j \cdot \vec\gamma_0
&= b_{j-1} - a_{j+1} \quad &&\text{for} \quad 2 \le j \le s\\
\vec\gamma_{s+1} \cdot \vec\gamma_0
&= b_s -1.
\endalignedat \tag19
$$
Of course, unless $s=1$, these conditions can not be solved with $\vec\alpha_s$
and $\vec\gamma_s$ that have only one component.  In general $(s > 1)$, these
variables should be multicomponent vectors, i.e. the number of free fields
involved (or the number of possible values for the indices $\mu$ and $\nu$ in
(1)) should be at least $s$ or $s+1$.  In fact, the vectors $\vec\gamma_j$ with
$1 \le j \le s+1$, can be chosen proportional to the simple roots of the Lie
algebra $\sel (s+2)$.  The points $0,1$ (and implicitly $\infty$) on the
Riemann sphere are obviously distinguished in the integral representation of
the hypergeometric functions.  It can be assumed that they are fixed by a
rational $\operatorname{SL}(2,\Bbb C)$ transformation --- a symmetry of the
Riemann sphere, otherwise, the formulas would give hypergeometric functions
with arguments of the form $(z-z_0)(z_1-z_\infty)/(z-z_\infty)(z_1-z_0)$.

With the extension of the above results to $q$--series in mind, it is practical
to rewrite eq. (18) in a slightly different form.  Let $\gamma_{ij} \equiv
\vec\gamma_i \cdot \vec\gamma_j$ and replace $V_{\vec\gamma_j}(t)$ in (18) by
$e^{i\hat\phi_j(t)}$ where the free fields $\hat\phi_j$ are the following
linear combinations of the fields $\phi^\mu \: \hat \phi_j(t) = \gamma_j^\mu
\phi^\mu(t)$.  It follows that
$$
\langle \hat\phi_i(t) \hat\phi_j(t^\prime)\rangle = \gamma_{ij}
\log(t-t^\prime).\tag20
$$
In order to obtain a representation of the hypergeometric functions which is
valid for real values of the argument $z$ between $0$ and $1$, it is enough to
require that (20) be true only for real $t$ and $t^\prime$ such that $t >
t^\prime$.  Then one has
$$
\multline
{}_{s+1} F_s(a_1, \dotsc , a_{s+1}; b_1, \dotsc , b_s; z = t_{s+1}) \cdot
\prod_{j=1}^s \widehat B(a_{j+1} b_j)\\
= \prod_{j=1}^s \int_0^{t_{j+1}}  dt_j \langle \prod_{j=1}^{s+2}
e^{i\hat\phi_j(t_j)}\rangle.
\endmultline \tag 21
$$

\head  3. Integral representations of the $q$--hypergeometric functions
\endhead

We now turn to the $q$--hypergeometric functions. We shall adopt the following
definition\footnote"$^\dag$"{Our definition of the $q$--hypergeometric
functions
which is not conventional is such that in the limit $q \rightarrow 1$, ${}_ r
\varphi_s(a_1,\dots,a_r;  b_1,\dots,b_s;  z) \rightarrow {}_rF_s(a_1,\dots,a_r;
b_1,\dots,b_s; z)$. It is close to the definition given in the second book of
ref.~\cite{1}. The presence of the factor $[(-1)^n  q^{n(n-1)/2}]^{1+s-r}$ is
to
ensure that a series of the form \thetag{22} is obtained when limits that
change the
difference $r-s$ are performed. (See the third book of ref.~\cite{1}). Note
that this
factor is absent when $r=s+1$.} for the $q$--analogs of the functions given in
\thetag{2}:
$$
\align
{}_r  \varphi_s(a_1,\dots,a_r & ;  b_1,\dots,b_s;  z)\\
&\equiv \sum^\alpha_{n=0} \ \frac{(a_1 \mid q)_n \dots (a_r \mid q)_n \cdot
z^n}
{(b_1 \mid q)_n \dots (b_s \mid q)_n (1 \mid q)_n} \ \left[ (-1)^n
q^{n(n-1)/2}
\right]^{1+s-r},
\tag22\\
\intertext{where}
(a \mid q)_n &\equiv \frac{\Gamma_q(a+n)}{\Gamma_q(a)} =
\frac{(q^a,q)_n}{(1-q)^n}
\equiv \prod^n_{k=1} \ \frac{1-q^{a+k-1}}{1-q},
\tag23
\endalign
$$
and in particular,\footnote"$^\dag$"{Note an amusing matrix-integral
representation
for this quantity \cite{23}:
$$
\frac{q^{n^2/2}}{(q,q)_n} = \frac{q^{n^2/2}}{\prod^n_{k=1}(1-q^k)} \sim \int
dH[dU]  e^{-m^2tr  H^2 + tr  HU  HU^\dag}.
$$
Here $H$ and $U$ are the Hermitean and unitary $n \times n$ matrices
respectively,
$[dU]$ denotes the Haar measure on $U(n)$, while $q=m^2 - \sqrt{m^4-1}$.
Interesting
implications of this representation are beyond the scope of the present
letter.}
$$
(1 \mid q)_n = \frac{(q,q)_n}{(1-q)^n} = \prod^n_{k=1} \ \frac{1-q^k}{1-q}.
$$
We are using here the standard notation for the $q$--shifted factorials:
$(z,q)_n
\equiv \prod^n_{k=1}(1-zq^{k-1})$. The $q$--gamma function $\Gamma_q$ is
defined so
that $\Gamma_q(z+1) = \frac{1-q^z}{1-q}  \Gamma_q(z),  \Gamma_q(1) = 1$.  In
the limit $q \rightarrow 1,  (a \mid q)_n \rightarrow (a)_n$.

The two ``elementary'' $q$--functions are:
$$
\align
{}_0\varphi_0(z) &= \sum^\infty_{n=0} \ \frac{q^{n(n-1)/2}}{(q,q)_n}  \bigl(
-(1-q)z \bigr)^n = \bigl( (1-q)z,q \bigr)_\infty\\
&\qquad \qquad \ \ \equiv E_q \left[ -(1-q)z \right] \equiv
\frac{1}{e_q[(1-q)z]}
; \tag24\\
{}_1\varphi_0(a,z) &\equiv \frac{1}{(1,z)^{[a]}} \equiv
\frac{(zq^a,q)_\infty}{(z,q)_\infty} = \prod^\infty_{k=1} \ \frac{1-zq^{k+a-1}}
{1-zq^k}.
\tag25
\endalign
$$

We shall make extensive use of the $q$--integral defined by
$$
\int^1_0  d_q t f(t) \equiv (1-q)  \sum^\infty_{n=0}  f(q^n)  q^n.
\tag26
$$
Some simple formulas involving this integral are:
$$
\align
\int^B_A  d_q t f(t) &= \int^1_0  d_q t  \left[ B f(Bt) - A f(At) \right];
\tag27\\
\int^1_{q^k}  d_q t f(t) &= (1-q)  \sum^\infty_{n=0}  \left[ f(q^n) - q^k
f(q^{n+k}) \right] = (1-q)  \sum^{k-1}_{n=0}  f(q^n)  q^n,
\tag28
\endalign
$$
provided $k$ is integer; in particular
$$
\int^1_q  d_q t  f(t) = (1-q)  f(1).
\tag29
$$
The function $\Gamma_q$ has the integral representation \cite{24}:
$$
\Gamma_q(z) = \int^{1/(1-q)}_0  d_q t \, t^{z-1}  E_q  [-q(1-q)t].
\tag30
$$
Especially useful will be the following representation for the $q$--beta
function:
$$
\aligned
\hat B_q(a,b) &\equiv B_q(a,b-a) \equiv \frac{\Gamma_q(a)  \Gamma_q(b-a)}
{\Gamma_q(b)}\\
&= \int^1_0  d_q t \, t^{a-1}(1,t_q)^{[b-a-1]} = \int^1_0  d_q t \, t^{a-1}
\frac{(t_q,q)_\infty}{(t_q^{b-a},q)_\infty}.
\endaligned
\tag31
$$

The three operations that are the $q$--generalizations of
\thetag{6}--\thetag{8} are
realized as follows. The linear transformation $t^n \longrightarrow
\frac{(a_{r+1} \mid q)_n}{(b_{s+1} \mid q)_n}$ which is required to effect
${}_r
 \varphi_s \rightarrow {}_{r+1}  \varphi_{s+1}$ is easily obtained from
\thetag{31}:
$$
\multline
\int^1_0  d_q t \, t^{a_{r+1}+n-1}  (1,tq)^{[b_{s+1}-a_{r+1}-1]}\\
= \frac{\Gamma_q(a_{r+1}+n)  \Gamma_q(b_{s+1}-a_{r+1})}{\Gamma_q(b_{s+1}+n)} =
\frac{(a_{r+1}\mid q)_n}{(b_{s+1} \mid q)_n}  \hat B_q(a_{r+1}, b_{s+1}).
\endmultline
\tag32
$$
It is then immediate to derive the following formula:
$$
\multline
{}_{r+1}  \varphi_{s+1}(a_1,\dots,a_{r+1};  b_1,\dots,b_{s+1};  z) =\\
\frac{1}{\hat B_q(a_{r+1},b_{s+1})} \int^1_0 d_q t \,  t^{a_{r+1}-1}
(1,qt)^{[b_{s+1}
-a_{r+1}-1]} {}_r  \varphi_s(a_1,\dots,a_r;  b_1,\dots,b_s;  tz)=\\
\frac{z^{-a_{r+1}}}{\hat B_q(a_{r+1},b_{s+1})} \int^z_0 d_q t \, t^{a_{r+1}-1}
(1,qt/z)^{[b_{s+1}-a_{r+1}-1]} {}_r  \varphi_s(a_1,\dots,a_r;  b_1,\dots,b_s;
t).
\endmultline
\tag33
$$
The realizations of the two remaining operations follow from the fact that
$$
\lim_{N\rightarrow\infty} \ \frac{\left( \frac{\log N}{\log q} \vert q
\right)_n}
{\bigl( N/(1-q) \bigr)^n} = (-1)^n  q^{n(n-1)/2}.
\tag34
$$
One then verifies that
$$
\aligned
{}_r  \varphi_s&(a_1,\dots,a_r;  b_1,\dots,b_s;  z) =\\
&\lim_{N\rightarrow\infty}  {}_{r+1}  \varphi_s \left( a_1,\dots,a_r,a_{r+1} =
\frac{\log N}{\log q};  b_1,\dots,b_s;  \frac{(1-q) z}{N} \right) =\\
&\lim_{N\rightarrow\infty}  {}_r  \varphi_{s+1} \left( a_1,\dots,a_r;
b_1,\dots,b_s,b_{s+1} = \frac{\log N}{\log q};  \frac{Nz}{1-q} \right).
\endaligned
\tag35
$$

\head 4. Free-field representation of the $q$--hypergeometric functions
\endhead

We now proceed to the free field interpretation of the above formulas.  As in
the case $q=1$, it will prove most natural for the functions ${}_{s+1}
\varphi_s$ (i.e. for those with $r-s =1$).  From the free field expression of
${}_{s+1} \varphi_s$, the corresponding representations of all the other
${}_r\varphi_s$ will be obtained by repeated use of the limits defined in (35).
 The main ingredient is of course the free field representation of the basic
function ${}_1\varphi_0(a;z) = 1/ (1,z)^{[a]} = (zq^a, q)_\infty /
(z,q)_\infty$.

To begin, we note that taking the ordinary logarithm of $(z,q)_\infty$ we get
$$
\aligned
\log (z,q)_\infty = \sum_{n=0}^\infty \log(1-zq^n)
&= \frac{1}{1-q} \int_0^1 \frac{d_qt}{t} \log(1-zt)\\
&= \frac{1}{1-q} \langle \phi(1) \int_0^1 \frac{d_qt}{t} \phi(zt) \rangle
\endaligned \tag36
$$
where $\phi(z)$ is the original free field that satisfies (1).  Now with the
help of (26) and (27) we arrive at
$$
\aligned
-\log \frac{(zq^\alpha, q)_\infty}{(z,q)_\infty}
&= \frac{1}{1-q} \langle \phi(1) \int_{q^\alpha}^1 \frac{d_qt}{t} \phi(zt)
\rangle\\
&= \frac{1}{(1-q)^2} \langle \int_q^1 \frac{d_qt}{t} \phi(t) \int_{q^\alpha}^1
\frac{d_qt}{t} \phi(zt)\rangle.\endaligned  \tag37
$$
This leads to the following definition of vertex operators:
$$
\align\split
V_\alpha(z,q)
&\equiv V_\alpha \{ \phi(z), q\} \equiv \frac{1}{1-q} \: \exp i
\int_{q^\alpha}^1 \frac{d_qt}{t} \phi(tz)\:\\
&\phantom{\equiv V_\alpha \{ \phi(z), q\}} = \frac{1}{1-q} \: \exp i\Phi_\alpha
(z)\: \endsplit\tag 38\\
V_{\vec\alpha}(z,q)
&\equiv \Pi_\mu V_{\alpha^\mu} \{\phi^\mu(z), q \}. \tag39
\endalign
$$
Equation (37) can now be interpreted as the statement that
$$
{}_1 \varphi_0(a;z) = \frac{1}{(1,z)^{[a]}} = \langle V_1(1,q) V_\alpha(z,q)
\rangle. \tag40
$$
Together with the integral representations of the previous section, this
relation allows us to represent any $q$--hypergeometric function in the form of
a correlator of free fields. (The screening charges $Q_{\vec\gamma,c} =
\int_0^c d_qt V_{\vec\gamma}(t,q)$ are now essentially double $q$--integrals.)

It is interesting to record the mode expansion of the free field $\Phi_\alpha
(z)$.  The free fields $\phi^\mu(z)$ on the Riemann sphere admit the expansion
$$
\phi^\mu(z) = - \sum_{n\ne0} \frac{a_n^\mu}{n} z^{_n} + a_0^\mu \log z + a^\mu
\tag 41
$$
where $a^\mu$ and $a_n^\mu$ $n \in \Bbb Z$ satisfy the commutation relations
$$
\aligned
[a_n^\mu, a_m^\nu]
&= \delta^{\mu\nu} \delta_{n+m,0} n\\
[a_0^\mu, a^\nu]
&= \delta^{\mu\nu}
\endaligned \tag42
$$
and generate the Heisenberg algebra.  This algebra has a Fock space
representation with the vacuum $\vert 0 \rangle$ defined by $a_n^\mu \vert 0
\rangle  = 0, n \ge 0$.  From (41) and (42) follows that $\langle \phi^\mu(z)
\phi^\nu(w) \rangle = \langle 0 \vert \phi^\mu(z) \phi^\nu(w) \vert 0\rangle =
\delta^{\mu\nu} \log (z-w)$.  From (38), we find that
$$
\Phi_\alpha(z) = (1-q) \sum_{n \ne 0} \frac{a_n^\mu}{n} z^{-n}
\frac{(1-q^{-\alpha n})}{(1 - q^{-n})} - a_0^\mu \log q^\alpha.
\tag43
$$

Now as was the case for $q = 1$, it is again convenient to introduce a new set
of massless free fields $\widehat \Phi_i(t)$ that satisfy here
$$
\langle \widehat\Phi_i(t) \widehat\Phi_j(t^\prime) \rangle \equiv \log
\left( t^{\gamma_{ij}} \left(1, \frac{qt^\prime}{t}\right)^{[\gamma_{ij}]}
\right) \quad \text{for real} \quad t > t^\prime \tag 44
$$
with $\gamma_{ij} = \vec\gamma_i \cdot \vec\gamma_j^\prime$ still the symmetric
matrix defined from the algebraic conditions (19).  In this case however, the
requirement that $t$ and $t^\prime$ are real and $t > t^\prime$ is less
trivial.  The point is that the expression on the r.h.s. of (44) is not
symmetric under the exchange of the pairs $i,t$ and $j,t^\prime$.  (This is in
contrast with the $q=1$ situations, see (1).)  One easily sees that the fields
$\widehat \Phi_i$ here have the following mode expansion
$$
\widehat\Phi_i(z) = -\sum_{n=0} \frac{\hat a_n^i}{n} z^{-n} + \hat a_0^i + \hat
a^i \tag 45
$$
with the operators $\hat a_n^i$, $n \in \Bbb Z$ and $\hat a^i$ satisfying the
commutation relations
$$
\aligned
[\hat a_n^i, \hat a_m^j]
&= -\delta_{n+m,0} q^{\vert n \vert(\gamma_{ij} + 1)/2}
\frac{[\gamma_{ij}n/2]}{[n/2]}\\
[\hat a_0^i, \hat a^j]
&= \gamma_{ij}.
\endaligned \tag46
$$
The symbol $[x]$ stands for $[x] = (q^x - q^{-x}) / (q-q^{-1})$.  In terms of
these $\widehat\Phi_i(t)$, the free field representation of the
$q$--hypergeometric functions is completely analogous to that of the functions
${}_rF_s$.  Explicitly, for $0< z< 1$, $z \in \Bbb R$,
$$
\multline
{}_{s+1} \varphi_s (a_1, \dotsc , a_{s+1}; b_1 , \dotsc , b_s; z = t_{s+1})
\prod_{j=1}^s \widehat B_q(a_{j+1}, b_j) =\\
z^{-a_{s+1}} \prod_{j=1}^s \left( \int_0^{t_{j+1}} d_q t_j \,
t_j^{a_{j+1}-a_j-1} \left( 1, \frac{qt_j}{t_{j+1}}\right)^{b_j-a_{j+1}-1}
\right) t_1^{a_1}(1,t_1)^{-[a_1]}\\
 = \prod_{j=1}^s \int_0^{t_{j+1}} d_qt_j \langle \prod_{j=0}^{s+2} e^{i
\widehat\Phi_j(t_j)} \rangle
\endmultline \tag 47
$$
For representations of the $q$--hypergeometric functions of an arbitrary
complex argument $z$, proper holomorphic analogs of the fields $\widehat
\Phi_i$ are required.

To sum up, we have described a surprisingly simple free field representation of
the $q$--hypergeometric functions.  It is related to the integral
representation of the functions ${}_{s+1} \varphi_s$ with the integrand given
as the product of $q$--powers of linear functions.

\remark{Acknowledgements}  We are grateful to A.H. Bougourzi, A. Leclair, V.
Spiridonov and R. Weston for inspiring discussions.  A.M. acknowledges the
hospitality of the Universit\'e de Montr\'eal.  The work of L.V. is supported
by grants from NSERC (Canada) and FCAR (Qu\'ebec).
\endremark

\enddocument